\documentclass[12pt]{article}
\usepackage{epsfig}
\usepackage{graphicx}
\usepackage{cite}
\usepackage{amsfonts}
\usepackage{amssymb}
\usepackage{latexsym}
\def\ee{\end{equation}}
\def\ba{\begin{eqnarray}}
\def\ea{\end{eqnarray}}

\def\bq{\begin{quote}}
\def\eq{\end{quote}}

 at 10truept

\newcommand{\beq}{\begin{equation}}
\newcommand{\eeq}{\end{equation}}
\newcommand{\beqa}{\begin{eqnarray}}
\newcommand{\eeqa}{\end{eqnarray}}
\newcommand{\bea}{\begin{eqnarray}}
\newcommand{\eea}{\end{eqnarray}}
\newcommand{\p}{\partial}
\newcommand{\al}{\alpha}
 
 \newcommand{\ep}{\epsilon}

\newcommand{\lmk}{\left(}
\newcommand{\rmk}{\right)}

\newcommand{\lle}{\left<}
\newcommand{\rgr}{\right>}
\newcommand{\lb}{\left|}
\newcommand{\rb}{\right|}

\newcommand{\overskrift}[1]{\vspace{6.0mm}\noindent\textbf{#1}\vspace{1.5mm}}

\def\lesssim{~\mbox{\raisebox{-.6ex}{$\stackrel{<}{\sim}$}}~}

\def\ltap{\ \raise.3ex\hbox{$<$\kern-.75em\lower1ex\hbox{$\sim$}}\ }
\def\gtap{\ \raise.3ex\hbox{$>$\kern-.75em\lower1ex\hbox{$\sim$}}\ }
\def\gl{\ \raise.5ex\hbox{$>$}\kern-.8em\lower.5ex\hbox{$<$}\ }
\def\roughly#1{\raise.3ex\hbox{$#1$\kern-.75em\lower1ex\hbox{$\sim$}}}

\begin{document}

\thispagestyle{empty}
%\begin{titlepage}
%%%%%%%%%%%%%%%%%%%%%%%%%%%%%%%%%%%%%%%%%%%%%%%%%%%%%%%%%%%%%%%%%%%%%%%%
%%%%
%\noindent
\begin{flushright}
{\tt hep-th/0612138}\\
December 2006
\end{flushright}

\vskip2cm
\begin{center}
{\Large{\bf On the One Loop Corrections to Inflation II:}}\\
{\Large{\bf The Consistency Relation}}\\
\vskip2cm {\large Martin S. Sloth\footnote{\tt
sloth@phys.au.dk}}\\

\vspace{.5cm}

\vskip 0.1in

{\em Department of Physics and Astronomy, University of Aarhus}\\ 
{\em DK-8000 Aarhus C, Denmark}\\

\vskip 0.1in
\vskip 0.1in
\vskip .25in
{\bf Abstract}
\end{center}
In this paper we extend our previous treatment of the one-loop corrections to inflation. Previously we calculated the one-loop corrections to the background and the two-point correlation function of inflaton fluctuations in a specific model of chaotic inflation. We showed that the loop corrections depend on the total number of e-foldings and estimated that the effect could be as large as a few percent in a $\lambda\phi^4$ model of chaotic inflation. In the present paper we generalize the calculations to general inflationary potentials. We find that effect can be as large as 70\% in the simplest model of chaotic inflation with a quadratic $m^2\phi^2$ inflationary potential. We discuss the physical interpretation of the effect in terms of the tensor-to-scalar consistency relation. Finally, we discuss the relation to the work of Weinberg on quantum contributions to cosmological correlators.  
%\end{abstract}

\vfill \setcounter{page}{0} \setcounter{footnote}{0}
\newpage
%\end{titlepage}

%%%%%%%%%%%%%%%%%%%%%%%%%%%%%%%%%%%%%%%%%%%%%%%%%%%%%%
%%%%%%%%%%%%%%%%%%%%%%%%%%%%%%%%%%%%%%%%%%%%%%%%%%%%%%
\setcounter{equation}{0} \setcounter{footnote}{0}
%%%%%%%%%%%%%%%%%%%%%%%%%%%%%%%%%%%%%%%%%%%%%%%%%%%%%%
%%%%%%%%%%%%%%%%%%%%%%%%%%%%%%%%%%%%%%%%%%%%%%%%%%%%%%

\section{Introduction}

From the point of view of inflationary model building, we are entering a very interesting era. While in the nineties a plethora of inflationary models was constructed, we are only now beginning to be able to discriminate between them experimentally. One of the important statements in the announcement of the WMAP 3 data set, was the claim that $\lambda\phi^4$ inflation is ruled out \cite{Spergel:2006hy,Kinney:2006qm,Tegmark:2006az}. Although this turns out not to be true if one allows for a non-vanishing neutrino fraction in the universe \cite{Hamann:2006pf}, it illustrates the precision with which WMAP and the upcoming PLANCK satellite can probe inflationary physics. As a consequence, we will soon need to take sub-leading effects into account in our theoretical predictions from various models.

The scope of this paper is to consider the one-loop effects in simple monomial models of chaotic inflation. If inflation has lasted only a short time, not much more than $65$ e-foldings, one can for most practical purposes treat the Hubble parameter, $H$, and the slow-roll parameters, $\ep$, $\eta$, as constant  during inflation. In this approximation one will find that the one-loop effects on the power spectrum are suppressed by a factor $H^2/M_p^2$, making them completely irrßelevant for observations. However, if inflation has lasted very long this approximation breaks down and there is an enhancement of the loop effects in the infrared. If $H_i$ is the initial Hubble parameter at the beginning of inflation and it is much larger than the Hubble parameter, $H_*$, when the physically observable modes exit the horizon during the last $65$ e-foldings or so, then  the loop effects can be enhanced by a factor $(H_i^2/H_*M_p)^{n}$,  where $n$ is some model dependent power \cite{Sloth:2006az}. This is due to the fact that the infrared cutoff on the loop-momentum is given by the initial size of the inflating patch, which is determined essentially by the initial value of the Hubble parameter.

In models of chaotic inflation \cite{Linde:1983gd},  where one typically has a very large total number of e-foldings of inflation, it turns out that the enhancement of the one-loop effects can be so large that it may be observationally relevant. The back-reaction of one-loop effects on the classical background has been calculated earlier in models of chaotic inflation, and shown to be large \cite{Abramo:1998hi,Abramo:2001dc,Abramo:2001dd,Mukhanov:1996ak,Abramo:1997hu,Afshordi:2000nr,Finelli:2001bn,Finelli:2003bp,Losic:2005vg,Martineau:2005aa,Wu:2006xp}, although it is not clear that this has any physical relevance in single field models, since the effect on the local expansion rate can be gauged away \cite{Unruh:1998ic,Geshnizjani:2002wp,Geshnizjani:2003cn}.  The type of infrared divergence that leads to the logarithmic growth of the one-loop correction with the scale factor, has  also been claimed to induce an effective mass for the photon during inflation in scalar quantum electrodynamics and thus leads to generation of magnetic fields \cite{Dimopoulos:2001wx,Prokopec:2004au,Prokopec:2002jn,Prokopec:2002uw,Prokopec:2003bx,Kahya:2006ui}. An other approach to loop effects, the stochastic approach, has been applied by Linde \cite{Linde:1986fd} (see also \cite{Goncharov:1987ir}), in order to understand the global structure of space-time in eternal inflation. 

Despite the significant amount of work on the back-reaction of one-loop effects on the classical background, the one-loop correction to the two-point quantum correlation function of inflaton field fluctuations was only recently evaluated explicitly in ref.~\cite{Sloth:2006az}  for the first time. The one-loop correction to the two point function is dominated by the seagull-diagram, and thus the calculation requires the fourth order action of the field perturbations, which was not available until it was calculated in the appendix of  ref.~\cite{Sloth:2006az}. In ref.~\cite{Sloth:2006az} it was furthermore estimated, that the one-loop correction to the two-point function of inflaton field perturbations in a model of $\lambda\phi^4$ chaotic inflation can be as large as a few percent. Subsequently, the fourth order action, with the inclusion of vector modes, was also given by Seery, Lidsey and Sloth \cite{Seery:2006vu}, who derived the primordial tri-spectrum of curvature perturbations.  Although the one-loop correction to the two-point correlation function of inflaton field fluctuations was not calculated explicitly before ref.~\cite{Sloth:2006az}, a general estimate of the effect of loop contributions on cosmological correlation functions appeared earlier in the work of Weinberg \cite{Weinberg:2005vy,Weinberg:2006ac}.

In the present paper we generalize the calculations of ref.~\cite{Sloth:2006az}, and investigate the effects in more details. We find that in simple models of chaotic inflation the one-loop effects can be very large and may have important effects on cosmological observables, such as the tensor-to-scalar ratio.

The paper is organized in the following way. In section 2, we take the super-horizon limit of the third order action, which was derived by  Maldacena \cite{Maldacena:2002vr}, and of the fourth order action given in ref.~\cite{Sloth:2006az,Seery:2006vu}. In section 3.1 we apply the super-horizon limit of the third order action to calculate the one-loop back-reaction on the classical background. In section 3.2 we calculate the one-loop correction to the two-point correlation function of inflaton fluctuations, using the super-horizon limit of the fourth order action. In section 3.3, we  discuss the physical implications in terms of the tensor-to-scalar consistency relation. In section 3.4, we discuss the relation to the work of Weinberg \cite{Weinberg:2005vy,Weinberg:2006ac}. Finally, in section 4 we roundup with a discussion.

\section{Effective action of perturbations}

The one-loop corrections to the two-point function of inflaton quantum fluctuations is dominated by the seagull diagram, which contains a vertex with four legs. Thus, in order to calculate it, we need the effective action of inflaton perturbations to fourth order in the field fluctuations.  Below we will review the calculation of the fourth order action and use it to calculate the effective interaction for a general inflaton potential in the super-horizon limit. 

\subsection{ADM formalism}

It is convenient to use the ADM formalism \cite{Arnowitt:1962hi}
to derive the action for the inflaton perturbations. Let us
consider the scalar action of the inflaton field
 \beq
S= \frac{1}{2}\int\sqrt{g}\left[R-(\p\phi)^2-2V(\phi)\right]~,
 \eeq
in the ADM metric, given by
 \beq
ds^2 =
-\mathcal{N}^2dt^2+h_{ij}(dx^i+\mathcal{N}^idt)(dx^j+\mathcal{N}^jdt)~.
 \eeq
In this metric, the action becomes \cite{Arnowitt:1962hi}
 \beq
S=\frac{1}{2}\int\sqrt{h}\left[\mathcal{N}R^{(3)}-2\mathcal{N}V+\mathcal{N}^{-1}\left(E_{ij}E^{ij}-E^2\right)+\mathcal{N}^{-1}
\left(\dot\phi-\mathcal{N}^i\p_i\phi\right)^2-\mathcal{N}h^{ij}\p_i\phi\p_j\phi\right]~,
 \eeq
where
 \beq
E_{ij}=\frac{1}{2}\left(\dot h_{ij}-\nabla_i
\mathcal{N}_j-\nabla_j \mathcal{N}_i\right)~.
 \eeq
We find it convenient to discuss
the effective action of the inflaton perturbations in the uniform
curvature gauge, where, when ignoring vector and tensor modes\footnote{In general one should also include vector and tensor contributions in the calculations, since they can seed scalar perturbations to higher order in perturbation theory. However, the contribution from the tensor modes tends to be suppressed \cite{Maldacena:2002vr} and the gradient structure of the vector modes \cite{Seery:2006vu} appears to exclude leading order IR divergent contributions of the kind we are looking for.}. The tensor modes are expected to be suppressed    , we have
 \beq
\phi = \phi_c+\delta\phi~,\qquad h_{ij}=a^2\delta_{ij}~,\qquad
\mathcal{N}=1+\al~,\qquad \mathcal{N}^i=\p_i\chi~.
 \eeq
In section 3.4, we will elaborate on the physical motivations for choosing this particular gauge for our calculations. 
 
The benefit of the ADM formalism is that the constraint
equations are easily obtained by varying the action in $N$ and
$N_i$, which acts as Lagrange multipliers. In this way the
constraint equations in the uniform curvature gauge become
 \beq
-a^2\delta^{ij}\p_i\phi\p_j\phi-2V-\mathcal{N}^{-2}\left(E_{ij}E^{ij}-E^2+\left(\dot\phi-\mathcal{N}^i\p_i\phi\right)^2\right)=0~,
 \eeq
and
 \beq
\nabla_j\left[\mathcal{N}^{-1}\left(E^i_j-\delta^i_jE\right)\right]
=
\mathcal{N}^{-1}\left(\dot\phi-\mathcal{N}^j\p_j\phi\right)\p_i\phi~.
 \eeq
If one perturbs the action by taking
 \beq
\phi=\phi_c+\delta\phi~,\qquad \al=\al_1+\al_2+\dots~,\qquad
\chi=\chi_1+\chi_2+\dots~,
 \eeq
and solves the constraint equations order by order, one finds to
first order \cite{Maldacena:2002vr}
 \beq \label{a1x1}
\al_1 = \frac{1}{2}\frac{\dot\phi_c}{H}\delta\phi~,\qquad
\p^2\chi_1=-\frac{1}{2}\frac{\dot\phi_c}{H}\dot{\delta\phi}-\frac{1}{2}\dot\phi_c\frac{\dot
H}{H^2}\delta\phi+\frac{1}{2}\frac{\ddot\phi}{H}\delta\phi~.
 \eeq
It is now trivial to obtain the second order action,
\bea
S_2 &=&\frac{1}{2}\int a^3\left[\delta\dot\phi^2-\p^i\delta\phi\p_i\delta\phi-V''\delta\phi^2+\frac{\dot\phi_c^2}{H^2}V\delta\phi^2\right.\nonumber\\
& &\qquad \left. +2\frac{\dot\phi_c^2}{H^2}\left(\frac{\ddot\phi_c^2}{H\dot\phi_c}+\frac{\dot\phi_c^2}{2H^2}\right)\delta\phi^2\right]~.
\eea

Generally, in order to obtain the action to order $n$, one
only needs to derive the constraint equations to order $n-1$, since the
$n$'th order terms multiplies the constraint equation to zero'th
order. In fact, it turns out in practice that the terms in the constraint equation to order $n-2$ cancels out as well. This implies that one only needs the first order terms
in eq.~(\ref{a1x1}) in order to obtain the action to
third order in perturbations. One obtains
 \bea \label{S3}
S_3 &=& \int
a^3\left[-\frac{1}{4}\frac{\dot\phi_c}{H}\dot{\delta\phi}^2\delta\phi-\frac{1}{4}\frac{\dot\phi_c}{H}\delta\phi(\p\delta\phi)^2-\dot{\delta\phi}\p^i\chi_1\p_i\delta\phi
\right.\nonumber\\
& & +\frac{3}{8}\frac{\dot\phi_c^3}{H}\delta\phi^3
-\frac{1}{4}\frac{\dot\phi_c}{H}V_{,\phi\phi}\delta\phi^3-\frac{1}{6}V_{,\phi\phi\phi}\delta\phi^3+\frac{1}{4}\frac{\dot\phi_c^3}{H^2}\delta\phi^2\dot{\delta\phi}
+\frac{1}{4}\frac{\dot\phi_c^2}{H}\delta\phi^2\p^2\chi_1\nonumber\\
&
&~\left.+\frac{1}{4}\frac{\dot\phi_c}{H}\left(-\delta\phi\p^i\p^j\chi_1\p_i\p_j\chi_1+\delta\phi\p^2\chi_1\p^2\chi_1\right)\right]~,
 \eea
as first derived by Maldacena \cite{Maldacena:2002vr}, and
subsequently generalized in
ref.~\cite{Creminelli:2003iq,Seery:2005wm,Seery:2005gb}. In eq.~(\ref{S3}), we have inserted the expression for $\al_1$ in order to bring it on the same form as in ref.~\cite{Maldacena:2002vr}. By going
one order further, one can in a similar fashion obtain the action
to fourth order in perturbations \cite{Sloth:2006az,Seery:2006vu}
 \bea \label{S4}
  S_4 = \int a^3 \left[ - \frac{1}{24} V_{,\phi\phi\phi\phi}
  \delta\phi^4 + \frac{1}{2} \partial_j \chi_1 \partial^j
  \delta\phi \partial_m \chi_1 \partial^m \delta\phi -
   \delta\dot{\phi} \partial_j \chi_2  \partial^j \delta\phi \right. \\\nonumber
   + (\alpha_1^2 \alpha_2 - \frac{1}{2}\alpha_2^2)
  (-6H^2 + \dot{\phi}^2) +
  \frac{\alpha_1}{2} \left\{ - \frac{1}{3} V_{,\phi\phi\phi}
  \delta\phi^3  -
  2\alpha_1^2 V_{,\phi} \delta\phi \right.\\ \nonumber
  + \alpha_1 \left( -  \partial^i
  \delta\phi \partial_i \delta\phi -
  V_{,\phi\phi} \delta\phi^2 \right) - 2 \partial_i \partial_j \chi_2
  \partial^i \partial^j \chi_1 + 2 \partial^2
  \chi_2 \partial^2 \chi_1  \\ \nonumber
\left. \left. + 2\dot{\phi}
  \partial_j \chi_2 \partial^j \delta\phi+
  2 \delta\dot{\phi}\partial_j \chi_1
  \partial^j \delta\phi \right\}\right] .
 \eea
We note that $\al_3$ has cancelled out of the action, so we only need the solution to the constraint equations to second
order \cite{Sloth:2006az,Seery:2006vu},
 \beq
\al_2=\frac{\dot\phi_c^2}{8H^2}\delta\phi^2+F(\delta\phi,\dot\phi)~,
 \eeq
and
 \bea
\p^2\chi_2 &=&
\frac{3}{8}\frac{\dot\phi_c^2}{H}\delta\phi^2+\frac{3}{4}\frac{\ddot\phi_c}{\dot\phi_c}\delta\phi^2-\frac{a^2}{4H}(\p\delta\phi)^2-\frac{1}{4H}\dot{\delta\phi}^2+\frac{\dot\phi_c}{2H}\p_i\chi_1\p_i\delta\phi\nonumber\\
& &+\frac{1}{4H}\left((\p^2\chi_1)^2-(\p_i\p_j\chi_1)^2\right)-\frac{V}{H}F(\delta\phi,\dot{\delta\phi})~,
 \eea
where we have for convenience defined \cite{Sloth:2006az}
 \beq \label{F}
F(\delta\phi,\delta\dot\phi)=
\frac{1}{2H}\p^{-2}\left[\p^2\al_1\p^2\chi_1-\p_i\p_j\al_1\p_i\p_j\chi_1+\p_i\dot{\delta\phi}
\p_i\delta\phi+\dot{\delta\phi}\p^2\delta\phi\right]~.
 \eeq

\subsection{The infrared limit}

The dominant one-loop contribution to the two-point correlation function of inflaton fluctuations comes from the infrared part of the seagull-diagram shown in fig.~(1). In order to evaluate it, we therefore only need the infrared part of the fourth order action, which is equivalent to the super-horizon limit of it.

To linear order in perturbations, the perturbation equation yields
 \beq
 \ddot{\delta\phi}+3H\dot{\delta\phi}-\frac{1}{a^2}\nabla^2\delta\phi + \left(V_{\phi\phi}-6\ep H^2\right)\delta\phi =0~.
 \eeq
Normalized to the Bunch-Davis vacuum in the infinite past, the solution for the mode function in Fourier space is given by
\beq
\delta\phi_k(\eta) =\frac{\sqrt{\pi}}{2}H\eta^{3/2}H_{\nu}^{(2)}(k\eta)~,
\eeq
where $\nu =3/2+3\ep-\eta$ and $\ep\equiv\dot\phi^2/2H^2$, $\eta\equiv V_{\phi\phi}/3H^2$ are the slow-roll parameters. In the super-horizon limit, where we can neglect the gradient terms, it becomes
\beq
\delta\phi \approx A H k^{-3/2}\left(\frac{k}{aH}\right)^{\eta-3\ep}~.
\eeq
For convenience we have defined $A\equiv e^{-i\pi/2}2^{\nu-3/2}\Gamma(\nu)/\Gamma(3/2)$. It is easy to see that in this limit, we have 
 \beq \label{superhorap1}
 \dot{\delta\phi}_k= -(\eta-2\ep)H\delta\phi+\mathcal{O}(\ep^2)~.
 \eeq

\begin{figure}[!hbtp] \label{fig01}
\begin{center}
\includegraphics[width=6cm]{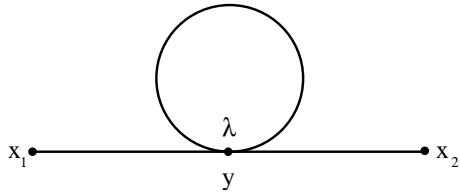}
\end{center}
\caption{The seagull diagram, which dominates the one-loop contribution to the two-point correlation function of inflaton field fluctuations. }
\end{figure}

Let us consider the interaction terms of the effective action in the super-horizon limit. By neglecting higher order gradient terms, the third order action $S_3$ in the super-horizon limit becomes
\bea
S_3^{SH}& =& \int
a^3\left[-\frac{1}{4}\frac{\dot\phi_c}{H}\dot{\delta\phi}^2\delta\phi-\dot{\delta\phi}\p_i\chi_1\p_i\delta\phi
 +\frac{3}{8}\frac{\dot\phi_c^3}{H}\delta\phi^3\right.\nonumber\\
& &\left.
-\frac{1}{4}\frac{\dot\phi_c}{H}V_{,\phi\phi}\delta\phi^3-\frac{1}{6}V_{,\phi\phi\phi}\delta\phi^3+\frac{1}{4}\frac{\dot\phi_c^3}{H^2}\delta\phi^2\dot{\delta\phi}
+\frac{1}{4}\frac{\dot\phi_c^2}{H}\delta\phi^2\p^2\chi_1\right]~.
\eea
To leading order in the slow-roll expansion, we can use the approximation
 \beq
 \p^2\chi_1\approx -\frac{1}{2}\frac{\dot\phi_c}{H}\dot{\delta\phi}+\mathcal{O}(\ep)~.
 \eeq
One has to be careful when estimating the slow-roll order of the terms involving $\dot{\delta\phi}$. Naively,  after partial integrations it appears that the order of the terms involving two factors of $\dot{\delta\phi}$ is lower than if we applied directly  eq.~(\ref{superhorap1}) in the action. One can show that, after a couple of partial space-time integrations, we would have up to a total time-derivative
\bea
\tilde S_3^{SH}&\approx&\int a^3 \left[-\frac{1}{2}\frac{\dot\phi_c}{H}H^2(3\eta-4\ep)\delta\phi^3-\frac{1}{6}V_{,\phi\phi\phi}\delta\phi^3-\frac{1}{2}\frac{\dot\phi_c}{H}(\p^{-2}\p_i\delta\phi)\p_i\delta\phi\left.\frac{\p\mathcal{L}}{\p\delta\phi}\right|_1\right]~.
\eea
The term proportional to the first order equation of motion can then be eliminated by a field redifinition
\beq
\delta\phi\to \delta\phi + \frac{1}{2}\frac{\dot\phi_c}{H}(\p^{-2}\p_i\delta\phi)\p_i\delta\phi~.
\eeq
However, the theory is not invariant under this point transformation of variables. In the transformation we have ignored a total time derivative term involving factors of the canonical conjugate field $\pi \propto   \dot{\delta\phi}$, and, as we show below in section 3.4, this is inconsistent when calculating quantum correlation functions. This implies that terms in the action which are non-linear in the canonically conjugate field cannot in general be simplified by partial time integrations, and the order of magnitude of their contribution is actually given by applying directly  eq.~(\ref{superhorap1}) in the action, when $\dot{\delta\phi}$ appears in the loop integral. When $\dot{\delta\phi}$ appears on an external leg, it does not introduce any further slow-roll suppression\footnote{We thank D. Seery for pointing this out.}. Thus, the correct expression for the super-horizon action to leading order in slow-roll is
\bea
S_3^{SH}&\approx&\int a^3 \left[-\frac{1}{4}\frac{\dot\phi_c}{H}H^2(3\eta-2\ep)\delta\phi^3-\frac{1}{6}V_{,\phi\phi\phi}\delta\phi^3 -\frac{1}{4}\frac{\dot\phi_c}{H}\dot{\delta\phi}^2\delta\phi\right.\nonumber \\ & &\left.\qquad+\frac{1}{2}\frac{\dot\phi_c}{H}\dot{\delta\phi}(\p^{-2}\p_i\dot{\delta\phi})\p_i\delta\phi\right]~.
\eea

In the same spirit, we can calculate the super-horizon limit of the $S_4$ action. Neglecting  the higher order gradient terms, the fourth order action reduces to
 \bea \label{S4sh}
S_4^{SH} &=& \int a^3\left[ -\frac{1}{24}V_{,\phi\phi\phi\phi}\delta\phi^4-\dot{\delta\phi}\p^i\chi_2\p_i\delta\phi\right.\nonumber\\
& &+\frac{1}{2}(2\al_1^2\al_2-\al_2^2)(-6H^2+\dot\phi_c^2)+\frac{1}{2}\al_1\left\{-\frac{1}{3}V_{,\phi\phi\phi}\delta\phi^3-2\al_1^2V_{,\phi}\delta\phi\right.\nonumber\\
& &\left.\left.+2(\p^2\chi_1\p^2\chi_2-\p_i\p_j\chi_1\p^i\p^j\chi_2)+2\dot{\phi}\p^i\chi_2\p_i\delta\phi -\al_1V_{,\phi\phi}\delta\phi^2\right\}
 \right]~.
 \eea
When evaluating the terms involving gradients one-by-one, we note that it is not possible to eliminate any time-derivative fields by time-like partial integrations, because the time-like partial integration will introduce new surface terms and terms involving higher order time derivatives of the perturbation fields, which can not be eliminated by a field redefinition without changing the perturbation theory. In this case, a $\dot{\delta\phi}$ term contributes effectively $\mathcal{O}(\ep\delta\phi)$ when it appears on an internal leg, as can be seen from eq.~(\ref{superhorap1}). In this way, we obtain to leading order in slow-roll in the super-horizon approximation that the important terms are
\bea
S_4^{SH}&\approx &\int a^3\left[-\frac{3}{8}\frac{\dot\phi_c^2}{H}\delta\phi^2\p^{-2}\p_j(\dot{\delta\phi}\p_j\delta\phi) +\frac{1}{8}\frac{\dot\phi_c^2}{H^2}\delta\phi^2\p_j(\dot{\delta\phi}\p^{-2}\p_j\dot{\delta\phi})\right.\nonumber\\
& &\qquad \left.+\frac{1}{16}\frac{\dot\phi_c^2}{H^2}\delta\phi^2\dot{\delta\phi}^2-\frac{1}{2}\dot{\delta\phi}^2F-VF^2\right]~,
\eea
where $F$ is given in eq.~(\ref{F}). To leading order it becomes
\beq
F\approx \frac{1}{2H}\p^{-2}\p^j(\dot{\delta\phi}\p_j\delta\phi)~.
\eeq

\section{One-loop corrections to inflation}

The Schwinger-Keldysh {\it real-time} formalism
\cite{Schwinger:1960qe,Keldysh:1964ud}  is appropriate for evaluating the one-loop corrections to expectation values self-consistently. It has also been extended to curved space and expanding backgrounds\cite{Calzetta:1986ey,Calzetta:1986cq,Boyanovsky:1992vi,Boyanovsky:1996rw,Boyanovsky:1997xt,Boyanovsky:2000hs} and used to study infrared divergences \cite{Boyanovsky:2004gq,Boyanovsky:2004ph,Boyanovsky:2005sh,Boyanovsky:2005px,Boyanovsky:2006kg}. For a review of the formalism, see the appendix of ref.~\cite{Collins:2005nu}. In this a approach the expectation value of some operator $\mathcal{O}$ is given by
\beq \label{exp1}
\lle 0\rb \mathcal{O} \lb 0\rgr = \frac{\lle 0\rb
T\left\{\mathcal{O}~e^{-i\int_{-\infty}^{0}d\eta\left[
H_I(\phi_c,\psi^+)- H_I(\phi_c,\psi^-)\right]}\right\}\lb
0\rgr}{\lle 0\rb T\left\{e^{-i\int_{-\infty}^{0}d\eta\left[
H_I(\phi_c,\psi^+)- H_I(\phi_c,\psi^-)\right]}\right\}\lb 0\rgr}~,
\eeq
if the initial state is the vacuum state $\lb 0\rgr$. A step function $\Theta(\eta-\eta_{infl})$ is absorbed in $H_I$, such that
the time integral effectively have $\eta_{infl}$ as lower limit.

This matrix element describes a system in the initial state
$\rho(\eta_{infl})$, evolved from conformal time $-\infty$ to $0$
with an operator inserted at $\eta$, and back again from $0$ to
$-\infty$, with a set of "+" fields on the increasing-time contour and a set of "-" fields on the
decreasing-time contour. The contractions between different pairs of the two types of
fields now yields four kinds of propagators
 \bea
 \lle 0\rb
T\left[\psi^{\pm}(x)\psi^{\pm}(x')\right]\lb 0\rgr & = & -i G^{\pm\pm}(x,x')\nonumber\\
&=&-i\int\frac{d^3k}{(2\pi)^3}e^{i\vec{k}\cdot(\vec{x}-\vec{x}')}G_k^{\pm\pm}(\eta,\eta')~.
 \eea
The time-ordering of the contractions then yields
 \bea
G^{++}_k(\eta,\eta')&=&
G^{>}_k(\eta,\eta')\Theta(\eta-\eta')+G^{<}_k(\eta,\eta')\Theta(\eta'-\eta)\nonumber\\
G^{--}_k(\eta,\eta')&=&
G^{>}_k(\eta,\eta')\Theta(\eta'-\eta)+G^{<}_k(\eta,\eta')\Theta(\eta-\eta')\nonumber\\
G^{-+}_k(\eta,\eta')&=& G^{>}_k(\eta,\eta')\nonumber\\
G^{+-}_k(\eta,\eta')&=& G^{<}_k(\eta,\eta')~,
 \eea
where
 \bea
G^{>}_k(\eta,\eta')&=& iU_k(\eta)U^*_k(\eta')\nonumber\\
G^{<}_k(\eta,\eta')&=& iU^*_k(\eta)U_k(\eta')~.
 \eea
One can of course also define $G^{>}(x,x')$, $G^{<}(x,x')$ from
which $G^{>}_k(\eta,\eta')$, $G^{<}_k(\eta,\eta')$  can obtained
by a Fourier transform.

From the previous section, it follows with $\psi\equiv\delta\phi$, that the effective interaction Hamiltonian in the super-horizon limit to leading order in slow-roll is 
 \bea
H_I(\phi_c,\psi^{\pm}) &\simeq& \int \frac{d^3y}{\eta^4
H^4}\left[\psi^{\pm}\lmk \ddot\phi_c+3H\dot\phi_c+V_{\phi}(\phi_c)\rmk +\lmk\frac{1}{6}V_{\phi\phi\phi}+\frac{1}{4}\frac{\phi_c'}{\mathcal{H}}H^2(3\eta-2\ep)\rmk{\psi^{\pm}}^3\right.\nonumber\\
& &\qquad +\left(-\frac{1}{4}\frac{\phi_c'}{\mathcal{H}}\frac{1}{a^2}{{\psi^{\pm}}'}^2\psi^{\pm}+
\frac{1}{2}\frac{\phi_c'}{\mathcal{H}}\frac{1}{a^2} {\psi^{\pm}}'(\p^{-2}\p_i{\psi^{\pm}}')\p_i\psi^{\pm} \right)
\nonumber\\
&  &
\qquad  +\left(\frac{1}{16}\frac{\phi_c'^2}{\mathcal{H}^2}\frac{1}{a^2}{{\psi^{\pm}}'}^2{\psi^{\pm}}^2+
\frac{1}{8}\frac{\phi_c'^2}{\mathcal{H}^2}\frac{1}{a^2} {\psi^{\pm}}^2\p_i({\psi^{\pm}}'\p^{-2}\p_i{\psi^{\pm}}') \right) \nonumber\\
 & &\left.\qquad  -{{\psi^{\pm}}'}^2F +VF^2 \right]\label{HI}
 ~,
 \eea
where $\mathcal{H}\equiv a'/a = aH$ and we have truncated the arguments of $\psi^{\pm}(\eta,y)$. The integration measure is given by the determinant of the de Sitter metric in conformal coordinates.

\subsection{One-loop corrections to the background}

The one-loop effective background equation of motion for the classical background field, $\phi_c$, follows directly from the tadpole renormalization condition. When we split the inflaton field in the classical background field, $\phi_c$, and the quantum fluctuation, $\delta\phi$, the tadpole condition defines what we mean by the classical background. The definition of the classical background field as the vacuum expectation value of the inflaton
\beq
\left< {}\right. \phi \left. {}\right> = \left< {}\right. \phi_c + \delta\phi \left. {}\right>  = \phi_c~,
\eeq
requires that the tadpole condition $\left< {}\right. \delta\phi \left. {}\right> = 0$ is satisfied to all orders. This implies that the one-loop effective background equation of motion is actually defined by the tadpole renormalization condition,  
\bea
0 &=& \lle \psi^{\pm}(x) \rgr_0 \nonumber\\
&=&-\int_{-\infty}^{\eta_0}d\eta\int\frac{d^3\vec{y}}{\eta^{4}H^4(\eta)}\left[\lmk
G^>(x,y)-G^<(x,y)\rmk\right.\nonumber\\
& \times&\left.\lmk
\phi''_c+2\mathcal{H}\phi_c'+V_{\phi} -i\lmk\frac{1}{2}V_{\phi\phi\phi}+\frac{3}{4}\frac{\phi_c'}{\mathcal{H}}H^2(3\eta-2\ep)\rmk
G^>(y,y)\rmk\right]~.
\eea
For this relation to be satisfied we must have
\beq
\ddot\phi_c+3H\dot\phi_c+V_{\phi} +\lmk\frac{1}{2}V_{\phi\phi\phi}+\frac{3}{4}\frac{\dot\phi_c}{H}H^2(3\eta-2\ep)\rmk \left< \delta\phi^2\right> =0~,
\eeq
where the last term on the left-hand-side is the one-loop correction to the tree-level background equation of motion. Any divergent part of this, or any piece that appears as a time-independent coupling, can be cancelled by counter terms in the effective action, but a finite time-dependent piece will generally be leftover from such a procedure and give rise to a small non-vanishing one-loop correction.

It is useful to consider a generic monomial type of inflation with the generic potential
\beq \label{pot1}
V(\phi) = \lambda M_p^{4-\al}\phi^{\al}~,
\eeq
such that the tree-level slow-roll parameters become
\beq
\ep =\frac{\al^2}{2}\frac{M_p^2}{\phi_c^2}~,\qquad \eta =\al(\al-1)\frac{M_p^2}{\phi_c^2}~.
\eeq
It is easy to verify that in the case $\al = 4$, the one-loop correction appears as an effective mass term \cite{Sloth:2006az}, but in general the form of the one-loop correction is non-trivial. In the slow-roll limit the one-loop effective equation of motion becomes
\beq \label{effeqmot}
\frac{\dot\phi_c}{H}\approx-\frac{V_{\phi}}{3H^2} -\lmk\frac{1}{2}\frac{V_{\phi\phi\phi}}{V}+\frac{1}{4}\frac{\dot\phi_c}{H}(3\eta-2\ep)\rmk \left< \delta\phi^2\right> ~.
\eeq
Using the definition $\ep_{eff}\equiv \dot\phi_c^2/(2H^2)$ of the one-loop effective slow-roll parameter  \cite{Sloth:2006az}, we obtain from eq.~(\ref{effeqmot})
\beq
\ep_{eff} \approx \ep + \delta\ep~,\qquad \delta\ep= \lmk\frac{V_{\phi}V_{\phi\phi\phi}}{2V^2}-\frac{1}{2}\ep(3\eta-2\ep)\rmk \frac{\left< \delta\phi^2\right>}{M_p^2} ~.
\eeq
The tree-level slow-roll parameters are evaluated at the time the observable modes exits the horizon, while the quantum two-point correlator $\left<\delta\phi^2\right>$ may contain information on the full history of inflation in a subtle way. In ref.~\cite{Sloth:2006az} we reviewed the evaluation of the two-point correlation function in details for the specific case of $\al= 4$. The generalization is given in \cite{Afshordi:2000nr}
\beq \label{dp2}
 \left< \delta\phi^2\right>= \frac{1}{12\al(4+\al)\pi^2}\lambda\left(\frac{\phi_c}{M_p}\right)^{2+\al}\left(\frac{\phi_i}{\phi_c}\right)^{4+\al}M_p^2~,
\eeq
where $\phi_i$ denotes the value of the classical field $\phi_c$ at the initial time $t_i$, which we can take to be the beginning of inflation. The initial value of the background inflaton field appears through the infrared cutoff on the loop-momentum. The infrared cutoff on the loop-momentum variable is given by $k_{IR}=a_iH_i$ and can be expressed in terms of $\phi_i$.

Using the slow-roll condition, we can write the total number of e-foldings of inflation as
\beq
N \equiv \ln\frac{a(t_*)}{a(t_i)} = \int_{t_i}^{t_*}H dt\approx \frac{1}{M_p^2}\int_{\phi_*}^{\phi_i}\frac{V}{V'}d\phi \approx \frac{1}{2\al M_p^2}\phi_i^2~,
\eeq
where we also applied the assumption $\phi_*<<\phi_i$. We conclude that $\left<\delta\phi^2\right>\propto N^{(4+\al)/2}$, which is consistent with the statement that the quantum correlator grows like a power of $\log a(t)$ \cite{Weinberg:2005vy} (we will return with a more detailed discussion of this in section 3.4). By considering the scenario of chaotic inflation and using the value $\phi_i$ at the end of the self-reproduction regime, and $\phi_c$ when the observable modes exit the horizon, we can estimate the largest possible loop correction. When the observable modes exit the horizon $N_* \simeq 60$ e-folding before the end of inflation, one can easily show that $\phi_*= \sqrt{2N_*\al}M_p$, while the field value at the end of the self-reproduction era is
\beq
\phi_i = \left(\frac{2\pi^2\al^2}{\lambda}\right)^{\frac{1}{\al+2}}M_p~.
\eeq
In order to match the observed level of CMB anisotropies we must further require $\lambda\approx 12\pi^210^{-10}(2N_*\al)^{-\al/2}$. Combining these theoretical and phenomenological constraints yields
\bea
\frac{\delta\ep}{\ep} &=& \left(\frac{V_{\phi\phi\phi}}{V_{\phi}}M_p^2-\frac{1}{2}(3\eta-2\ep)\right)\frac{ \left< \delta\phi^2\right>}{M_p^2}\nonumber\\
&\lesssim& \frac{10^{-10}}{2\al}\frac{4-3\al}{4+\al}\left(\frac{\al^2}{6\cdot 10^{-10}}\right)^{\frac{\al+4}{\al+2}}(2N_*\al)^{-\frac{2\al+8}{2\al+4}}~,  \label{de}
 \eea
where we have inserted $\phi_c$ for $\phi_*$. In fig.~(2), we have plotted the relative one-loop correction $\delta\ep/\ep$ to the slow-roll parameter $\ep$ for $N_*=45$ and $N_*=60$. The effect is a few percent for  $\al = 4$. For $\al =2$, the effect is order $10\%$ to $15\%$.

\begin{figure}[!hbtp] \label{fig01}
\begin{center}
\includegraphics[width=8cm]{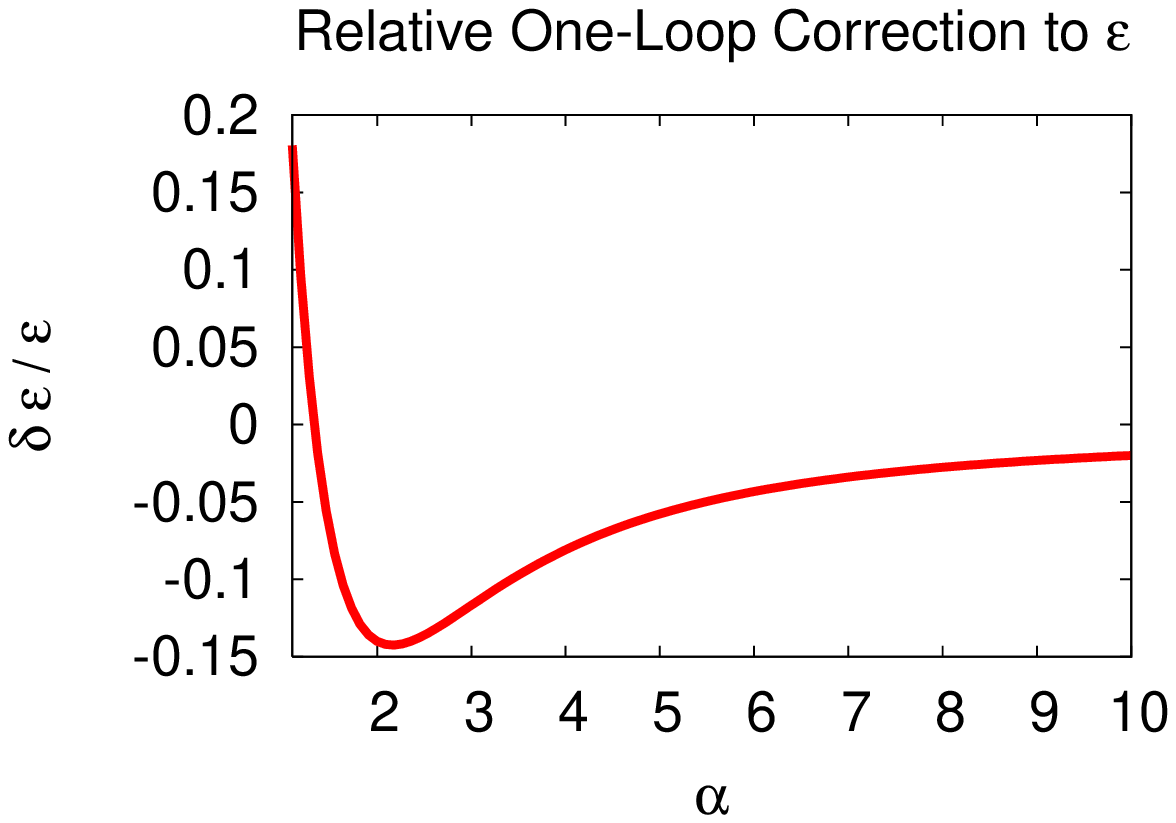}
\includegraphics[width=8cm]{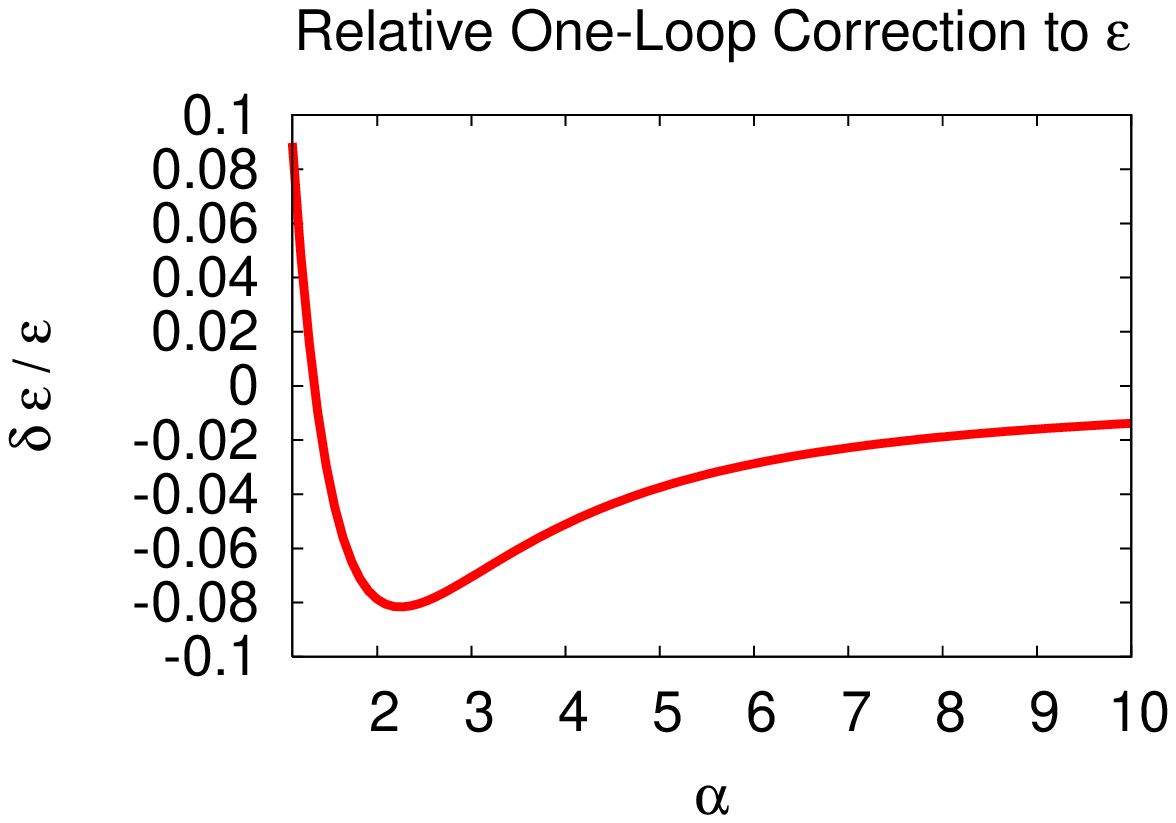}
\end{center}
\caption{The relative one-loop correction $\delta\ep/\ep$ to the slow-roll parameter $\ep$. In the left panel we have plottet the one-loop correction in the case of a low reheating temperature for $N_*=45$. In the right panel we have shown the same plot, but now with  $N_*=60$.}
\end{figure}

\subsection{One-loop corrections to the two-point function}

The two-point function, $T(\eta_0,k)\equiv\left<\psi_k^{\pm}(\eta_0)\psi_k^{\pm}(\eta_0)\right>$, evaluated to one-loop order can be organized in terms
of contributions to zero'th $T^{(0)}$, first $T^{(1)}$, and second
$T^{(2)}$  order in $\lambda$, where lambda is a generic coupling constant counting the number of vertices. Thus, the tree-level contribution is given by $T^{(0)}$, and to one-loop order we have $T =  T^{(0)}+T^{(1)}+T^{(2)}$. By using eq.~(\ref{HI}) in eq.~(\ref{exp1}), with four copies of $\psi^+$ inserted as the operator and doing the appropriate contractions, we can compute $T$. The leading one-loop contribution to the two-point function comes from the seagull diagram in fig.~(1).  From the seagull diagram, we obtain the following contribution to the two-point correlation function of inflaton fluctuations in Fourier space
 \bea\label{T11}
T^{(1)}(\eta_0,k)& =& -2
\int_{\eta_{infl}}^{\eta_0}\frac{d\eta}{\eta^2H^2}\textrm{Im}\left[\p_{\eta}G^{>}_k(\eta_0,\eta)\p_{\eta}G^{>}_k(\eta_0,\eta)\right] \nonumber\\
& &\qquad\times\left(\frac{1}{8}\ep+\frac{1}{4}(2\ep-\eta)\right)\int\frac{d^3k'}{(2\pi)^3}(-i)G^{>}_{k'}(\eta,\eta)\nonumber\\
& &-2
\int_{\eta_{infl}}^{\eta_0}\frac{d\eta}{\eta^3H^3}\textrm{Im}\left[\p_{\eta}G^{>}_k(\eta_0,\eta)G^{>}_k(\eta_0,\eta)\right] \nonumber\\
& &\qquad\times\frac{3}{16}(2\ep-\eta)H\int\frac{d^3k'}{(2\pi)^3}(-i)G^{>}_{k'}(\eta,\eta)~.
 \eea
 The time that appears in the lower limit of the integral is the end of the self-reproduction regime, but for most practical purposes we can take $\eta_{infl}\to-\infty$.  
 We have also used the approximation where momenta integrated over in the loop is much smaller than the external momentum  $k'<<k$, such that  $\p^{-2}\p^i(\psi'\p_i\psi)$ in Fourier space becomes $|{\bf k}+{\bf k}'|^{-2}(k^i+k'^i)(\psi'k'_i\psi)\simeq k'/k\psi'\psi$, if the conjugate field appears as an external leg and the other as an internal, which yields a loop integral that is not divergent in the infrared and can be ignored. Similarly if they both appear internal or external, the contribution becomes $(1/2) \psi'\psi$, due to momentum conservation at the vertex. By subsequently applying eq.~(\ref{superhorap1}) when the conjugate field appears on an internal leg, this leads to the last term in eq.~(\ref{T11}) above.

Now, we can follow the calculation of \cite{Sloth:2006az}. In terms of the mode functions $U_k(\eta)$, we have 
 \bea \label{T12}
T^{(1)}(\eta_0,k)& =& 2
\int_{\eta_{infl}}^{\eta_0}\frac{d\eta}{\eta^2H^2}\textrm{Im}\left[(\p_{\eta}U_k(\eta))^2U_k^{*2}(\eta_0)\right] \left(\frac{1}{8}\ep+\frac{1}{4}(2\ep-\eta)\right)\left< \delta\phi^2\right>\nonumber\\
& &+2
\int_{\eta_{infl}}^{\eta_0}\frac{d\eta}{\eta^3H^3}\textrm{Im}\left[ (\p_{\eta}U_k(\eta))U_k(\eta)U_k^{*2}(\eta_0)\right] \frac{3}{16}(2\ep-\eta)H\left< \delta\phi^2\right>~,
 \eea
where $\left< \delta\phi^2\right>$ is given in eq.~(\ref{dp2}). For the physically observable modes, which has spent only short time out side the horizon, in eq.~(\ref{T12}) it is a good approximation to assume \cite{Sloth:2006az}
\beq
U_k(\eta) =\frac{iH}{k\sqrt{2k}}(1+ik\eta)e^{-ik\eta}~.
\eeq
The conformal time integrals in eq.~(\ref{T12}) turn out to get their dominant contributions from the integration from $\eta_*$ to $\eta_0$, where $\eta_*$ is the conformal time at which the physically observable comoving momenta $k$ crosses outside the horizon. In the interval $[\eta_*,\eta_0]$, we can treat the two-point correlation $\left< \delta\phi^2\right>$, the slow-roll parameters $\ep$, $\eta$ and the potential $V$ and its field derivatives as constant. The integral then simplifies, and can easily be evaluated analytically. In the limit $x_0=-k\eta_0\to 0$ we obtain
\beq
\int_{\eta_{infl}}^{\eta_0}\frac{d\eta}{\eta^2H^2}\textrm{Im}\left[ (\p_{\eta}U_k(\eta))^2U^{*2}_k(\eta_0)\right]\approx  -\frac{H^2}{8k^3}~,
\eeq
and 
\beq
\int_{\eta_{infl}}^{\eta_0}\frac{d\eta}{\eta^3H^3}\textrm{Im}\left[ (\p_{\eta}U_k(\eta))U_k(\eta)U^{*2}_k(\eta_0)\right]\approx  \frac{H}{4k^3}\left[\textrm{Ci}(-2k\eta_0)-2\right]~.
\eeq
We finally obtain on super-horizon scales, the following one-loop corrected two-point function
\bea \label{dP}
\mathcal{P}(\eta_0,k)& \approx &
\frac{H^2}{4\pi^2}\left[1-\lmk \frac{1}{16}\ep+\frac{1}{2}(2\ep-\eta)-\frac{3}{8}(2\ep-\eta) \textrm{Ci}(-2k\eta_0)\rmk\left< \delta\phi^2\right>\right]~,
 \eea
which can be regarded as the generalization of eq.~(63) in ref.~\cite{Sloth:2006az}. The expression in  ref.~\cite{Sloth:2006az} is however one order higher in the slow-roll expansion, because we used the approximation in eq.~(\ref{superhorap1}) also when $\dot\delta\phi_k$ appeared in an external leg. It is reassuring that the corrections in eq.~(\ref{dP}) and eq.~(\ref{de}) are of the same order in the slow-roll expansion.

 \begin{figure}[!hbtp] \label{fig01}
\begin{center}
\includegraphics[width=8cm]{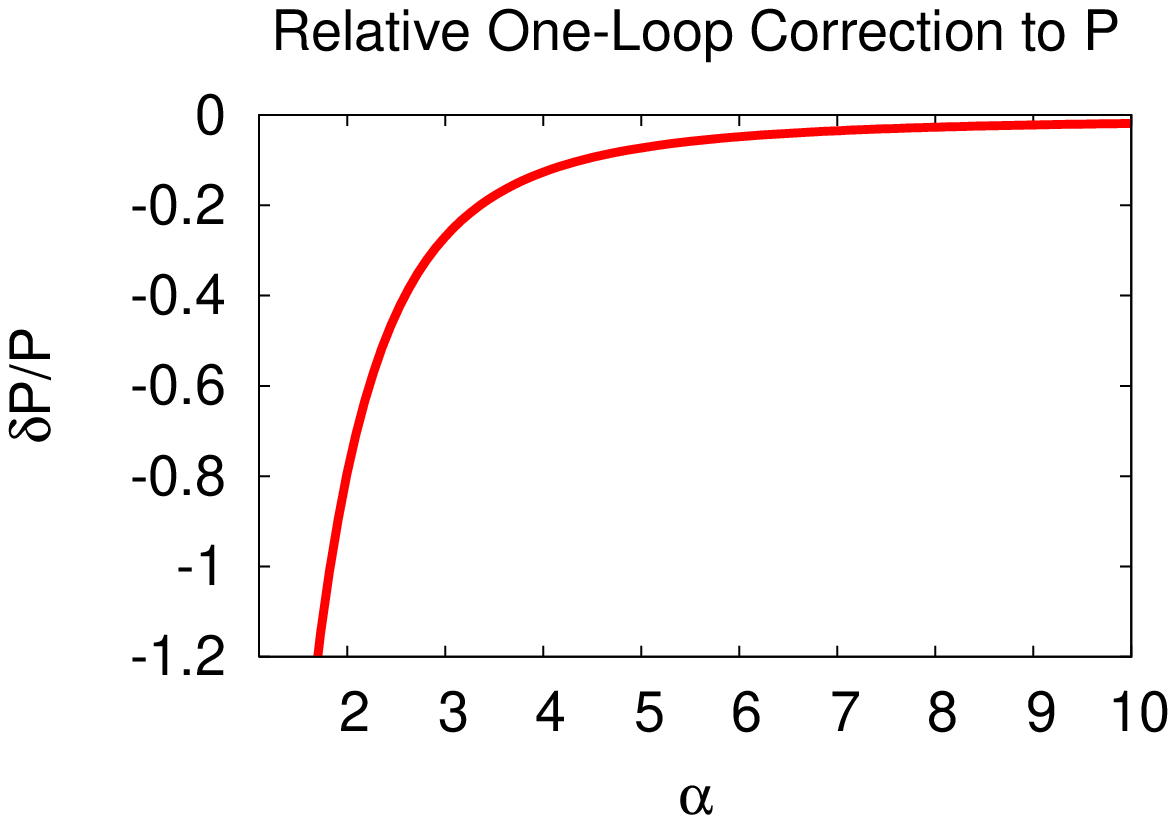}
\includegraphics[width=8cm]{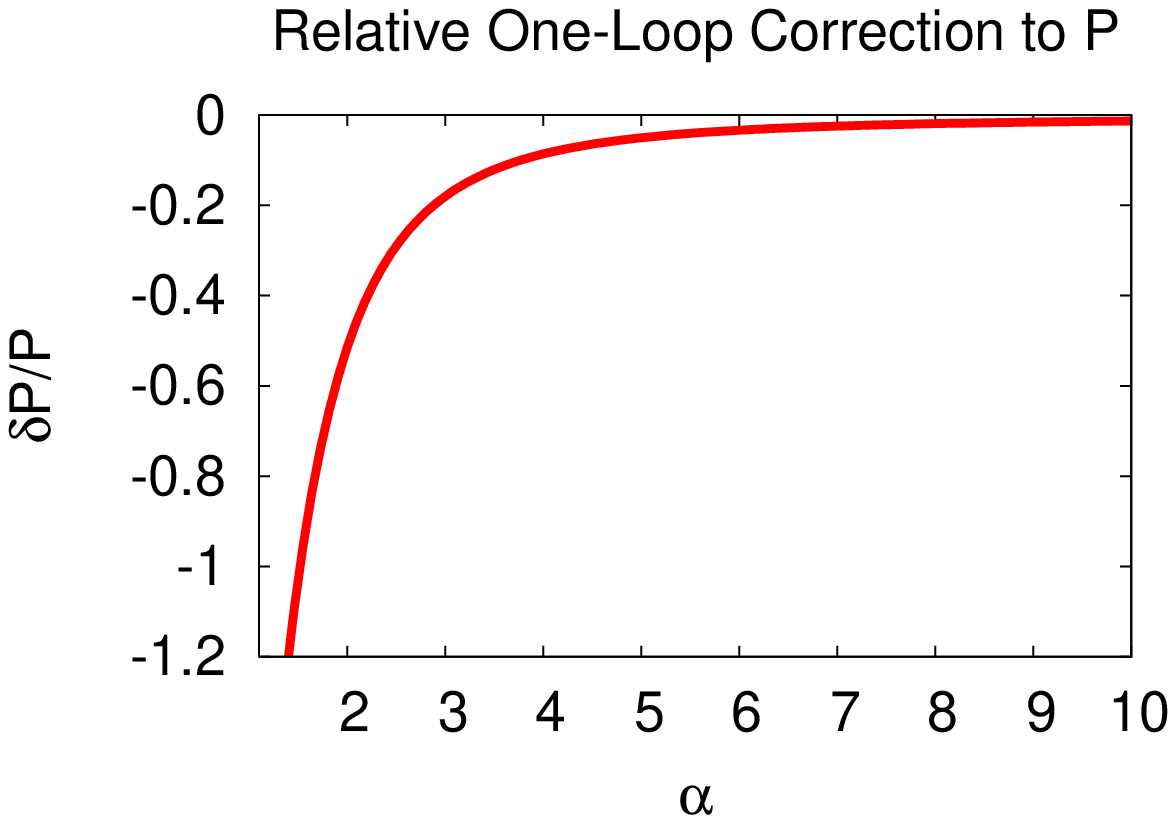}
\end{center}
\caption{The relative one-loop correction $\delta \mathcal{P}/\mathcal{P}$ to the power spectrum of inflaton fluctuations. In the left panel we have plottet the one-loop correction for $N_*=45$. In the right panel we have shown the same plot, but with $N_*=60$. We see that in the present approximation the maximal one-loop correction is of the order $5\%$ to $15\%$ for $\lambda\phi^4$, while it is $50\% $ to $70\%$ for a typical model of $m^2\phi^2$ chaotic inflation.}
\end{figure}

In fig.~(3), we have plotted the maximal relative one-loop correction to the power-spectrum  $\delta\mathcal{P}/\mathcal{P}$, by considering again the scenario of chaotic inflation with the potential given in eq.~(\ref{pot1}). Especially, we remark that for a model of chaotic inflation with a potential of the type $m^2\phi^2$, the one-loop corrections significantly influences our predictions for the two-point function of inflaton fluctuations. The effect appears to be of the order of 50\% with $N_*=60$ and as large as 70\% if $N_*=45$.

\subsection{Physical interpretation and the consistency relation}

With the WMAP data, it has been possible to start to severely constrain various inflationary models. In fact, it was claimed with the release of the WMAP 3 data, that the simple $\lambda\phi^4$ inflationary model is ruled out. It has later turned out not to be true, and $\lambda\phi^4$ is marginally allowed if the universe is composed with a non-vanishing neutrino fraction \cite{Hamann:2006pf}. This illustrates how sensitive data is becoming, to the exact theoretical predictions from various models of inflation.  

Crucial for ruling out different polynomial models of chaotic inflation is the predicted tensor-to-scalar ratio. Taking the chaotic model of  $\lambda\phi^4$ or $m^2\phi^2$ inflation seriously, we must ask for its precise prediction for the tensor-to-scalar ratio including loop effects, before we can even start to constrain it with data. In the previous sections we have seen that the loop effects can be rather significant. On the other hand, one could turn the argument around and view the one-loop corrections from a low-energy effective point of view, and claim that the results just imply that a pure $\lambda\phi^4$ model is not likely in the effective framework of chaotic inflation. However, the low-energy effective potential will have to be extremely finely tuned, if it is constructed such that it exactly mimics the loop effects. In other words, the $\lambda\phi^4$ model, or other monomial chaotic inflationary models, are very simple and well defined theoretical models when loop effects are included, and it is important to experimentally constrain them if possible.

At present, we have not yet calculated the loop corrections to the tensor perturbations. Before we can do this, we must have the fourth order action for tensor perturbations, which has not yet been calculated. Assuming that the loop corrections to the tensor perturbations are of same order of magnitude as the loop corrections to the scalar perturbations or smaller, and does not  exactly cancel the loop corrections to the scalar power spectrum when the tensor-to-scalar ratio is calculated, we can still estimate the size of the loop-correction. From the left panel of fig.~(3), we estimate that the correction to the tensor-to-scalar ratio is as large as $70\%$ for the simple $m^2\phi^2$ chaotic inflation model, while it is only a few percent for $\lambda\phi^4$. This is consistent with the result of ref.~\cite{Sloth:2006az}.

\begin{figure}[!hbtp] \label{fig01}
\begin{center}
\includegraphics[width=7cm]{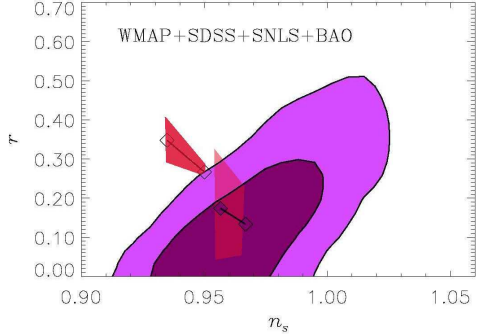}
\end{center}
\caption{The likelihood contours of the tensor-to-scalar ratio vs. the scalar spectral index \cite{Hamann:2006pf}. Two models are indicated. On the edge of the $95\% $ exclusion likelihood contour is the predictions from the $\lambda\phi^4$ model while in the middle of the $68\% $ exclusion likelihood contour the predictions of the $m^2\phi^2$ model is indicated. We have indicated the model predictions with $45<N_*<60$. If we did not take into account one-loop corrections, the predictions would be line-shaped between the squares. The full polygons indicates qualitatively the theoretical uncertainty when the one-loop correction to the two-point correlation function of inflaton fluctuations are included. }
\end{figure}

When constraining models of monomial chaotic inflation, it means the the predicted theoretical model lines in the spectral index, $n_s$, vs. tensor-to-scalar ratio, $r$, plots are no-longer line shaped but becomes blurred over a small region due to the loop effects. We have illustrated this in fig.~(4), where the small polygon shaped regions would have been line shaped, if we did not take into account  the one-loop correction to the two-point correlation function of inflaton fluctuations. We see that this can turn out to be important in the future, making it possible to discriminate between a model of $m^2\phi^2$ inflation  with a minimum number of total e-foldings of inflation and a model of  $m^2\phi^2$ chaotic inflation, with a huge total number of e-foldings. We find that this is an interesting possibility that deserves further investigation.

\subsection{Quantum contributions to general cosmological correlations}

The analysis carried forward in ref.~\cite{Sloth:2006az} and in the present work, is closely related to a slightly more general question lately addressed  by Weinberg \cite{Weinberg:2005vy,Weinberg:2006ac}. At tree-level it is well known that quantum correlations in cosmology only depend on the behavior of the unperturbed background field near the time of horizon exit. Weinberg asked wether this is still true when loop effects are included, or if the contribution of loop graphs can depend on the whole history of the unperturbed universe. It is was found that in general the correlations can depend on the whole history of the universe, although at most to powers of the logarithm of the scale factor \cite{Weinberg:2006ac}. As we will discuss in this subsection, this is consistent with our findings  \cite{Sloth:2006az}.

In \cite{Weinberg:2005vy,Weinberg:2006ac}, the expectation value of any product of operators in eq.~(\ref{exp1}) is expanded on the form 
\beq
\left<\mathcal{O}(t)\right> = \sum_{N=0}^{\infty} i^N \int_{-\infty}^t dt_N \int_{-\infty}^{t_N} dt_{N-1}\dots \int_{-\infty}^{t_2} dt_1
\left<\left[H_I(t_1),\left[H_I(t_2),\dots \left[H_I(t_N),\mathcal{O}(t)\right]\right]\right]\right>~.
\eeq
Until now we have consistently worked in the zero-curvature gauge, with the scalar perturbations given by Sasaki-Muhkanov variable, which is identical to the field fluctuations themselves in this particular gauge. In \cite{Weinberg:2005vy,Weinberg:2006ac} the comoving curvature gauge was chosen. In the comoving curvature gauge the field fluctuation vanishes and the instead the scalar perturbations are given by the comoving curvature perturbation $\zeta$, which is defined as the perturbation of the spatial section of the metric
\beq
g_{ij} = a^2e^{2\zeta}\delta_{ij}~,
\eeq
where we for simplicity are ignoring tensor perturbations and only consider scalar perturbations of the metric in single field inflation. In the interaction picture, the curvature perturbation field is
\beq\label{exp2}
\zeta({\bf x},t)= \int \frac{d^3k}{(2\pi)^{3/2}} \left[ \zeta_k(t)e^{i{\bf k\cdot x}}a_{\bf k}+\zeta^*_k(t)e^{-i{\bf k\cdot x}}a^{\dagger}_{\bf k}\right]~.
\eeq
In single field inflation $\zeta_k$ is conserved on super-horizon scales, because on these scales $\dot\zeta_k$ vanishes very fast as $a^{-2}$. So if $\zeta^0_k$ is the time independent limit of $\zeta_k$, the difference $\zeta_k-\zeta^0_k$ goes again essentially like $a^{-2}$. The crucial observation in \cite{Weinberg:2005vy,Weinberg:2006ac} is that this implies that the commutator of any two combination of $\zeta$-fields always goes as $a^{-3}$.

One can now evaluate the time dependence of any correlation function of $\zeta$'s  outside the horizon by observing that the Lagrangian density maximally carries three powers of the scale factor. Since in eq.~(\ref{exp2}) there are just as many commutators as there are interactions, and each commutator carries three powers of the inverse scale factor, there can at most be zero factors of $a(t)$ in any of the integrals over time in eq.~(\ref{exp2}). This implies that the integrands can at most grow like a power of $t$, which is similar to a power of $\ln a(t)$  \cite{Weinberg:2005vy,Weinberg:2006ac}. It is thus concluded that the quantum correlation function can at most grow like a power of $\ln a(t)$, and without huge amount of e-foldings the loop corrections can never become large.

This conclusion is in agreement with the results of  ref.~\cite{Sloth:2006az}, which indeed did find large loop corrections to the quantum correlations in chaotic inflation only with a very long period of total inflation. In fact it was found that the correction grows like a power of the total number of e-foldings $N$, which is equivalent to a power of $\ln a(t)$ (see for instance eq.~(29) of  \cite{Sloth:2006az}). 

However, there is a fundamental difference between the two approaches. In one approach it is the field perturbation (which is the Sasaki-Mukhanov variable in that gauge) which is quantized, while in the other approach it is the curvature perturbation which is quantized. The two approaches does not appear to be equivalent at the quantum level, since the transformation between the two variables is a non-trivial point transformation. To see this, consider the general subclass of canonical transformations given by point transformations, which transforms the canonical conjugated field, $\pi$, in the following way \cite{Matacz:1992tp}
\beq \label{eqpi}
\pi=\frac{\p{\mathcal{L}}}{\p\dot\psi}\to \pi-\frac{\p F[\psi,t]}{\p\psi}~,
\eeq
where the generating function for the transformation, $F$, appears as a total time derivative in the Lagrangian after the transformation, i.e.
\beq
\mathcal{L}[\psi,\dot\psi]\to \mathcal{L}[\psi,\dot\psi] -\frac{d}{dt}F[\psi,t]~.
\eeq
This implies that the action transforms as 
\beq
S[\psi]\to S[\psi]-F[\psi_i,t_i]+F[\psi_f,t_f]~,
\eeq
and does clearly not change the classical tree-level equation of motion, which is derived from requiring that the variation of the action in the field vanishes. On the other hand, the evolution operator and the states transforms as 
\beq
U(t_f,t_i)\to e^{-iF(\psi_f,t_f)}  U(t_f,t_i)e^{iF(\psi_i,t_i)}~,\qquad \left<\Psi(t)\right|\to   \left<\Psi(t)\right| e^{iF(\psi,t)} ~,
\eeq
such that, if we consider the expectation value of some composite operator $\mathcal{O}(\psi,\pi) $ in the transformed vacuum state, it yields
\beq
\left<\mathcal{O}(t)\right> = \left< U^{\dagger}(t,t_i)e^{iF(\psi,t)}\mathcal{O}(t)e^{-iF(\psi,t)}U(t,t_i)\right>~.
\eeq
Now, if $\mathcal{O}$ is only a function of $\psi$ everything commutes and the phases cancel out, and when $\mathcal{O}$ depends on $\pi$, one can use eq.~(\ref{eqpi}) to see that the phases will cancel out in general \cite{Matacz:1992tp}.  However, the same argument does not apply if $F$ depends on $\pi$, in which case the theory is not invariant beyond tree-level. In fact, if $F$ is linear in $\pi$, the transformation is still a point transformation, which will take $\psi \to \psi + \p F/\p\pi$. This is exactly the kind of transformation needed to transform from the $\zeta$ variable to the $\delta\phi$ variable on super-horizon scales, up to a trivial time-dependent scale transformation \cite{Maldacena:2002vr}. We may note, that when we are evaluating the $S$-matrix, we take the expectation value in a sandwich of states in the past and future infinity. This implies that the $S$-matrix is still invariant under the type of point transformation with $F$ linear in $\pi$, provided the field, $\psi$, vanishes fast enough in future and past infinity \cite{chisholm,Kamefuchi:1961sb}.

The tree-level correlation functions to $n$'th order are still independent of the phase $F$ on super-horzon scales, since, as discussed above, on those scales the commutator vanishes very fast as $a^{-3}$ and the fields essentially becomes classical  \cite{Weinberg:2005vy,Weinberg:2006ac}. However, on sub-horizon scales or for loop corrections where quantum effects becomes important, the two variables does not appear to yield equivalent results. In fact, this seems to raise a question regarding the physical interpretation during inflation of the non-linear quantity, constructed to be conserved on all scales in ref.~\cite{Enqvist:2006fs}.

One could argue for calculating the loop corrections in terms of  $\zeta$, since it is really the observable quantity, which is conserved on super-Hubble scales \cite{Weinberg:2005vy}. However, it is not clear that the classically conserved $\zeta$ is also conserved at the quantum level \cite{Weinberg:2005vy}. This imply that we would first have to calculate the loop corrections in order to find the conserved $\zeta$ in the one-loop effective theory. In addition, this argument would only be strictly valid in single field models, since in multi-field inflationary models $\zeta$ is anyway not conserved on super-Hubble scales.  On the other hand, the Sasaki-Mukhanov variable really represents the fundamental degrees of freedom, which appears to be the fluctuations of the matter fields. In terms of the matter field fluctuations, the splitting between the background and the fluctuations is very clear and well defined in terms of the tadpole renormalization condition. It thus appears to be more appropriate to calculate the loop corrections in terms of the Sasaki-Mukhanov variable.

\section{Discussion}

We have generalized and expanded the analysis  in ref.~\cite{Sloth:2006az} of the one-loop correction to the two-point correlation function of inflaton fluctuations. It is shown that in chaotic inflation, where inflation starts just below the self-reproduction regime and gives rise to huge total amount of e-foldings, the one-loop corrections may be physically significant. For the model of $\lambda\phi^4$ chaotic inflation investigated in ref.~\cite{Sloth:2006az}, our results are consistent with a maximum correction of a few percent if the observable modes exited the horizon about $60$ e-foldings before the end of inflation ($N_*=60$)  \cite{Sloth:2006az}. The effect can be slightly larger, of the order $15\% $, if the modes exited  $45$ e-foldings before the end of inflation ($N_*=45$). Having generalized our result to general inflationary potential and especially to general models of monomial chaotic inflation, we find it very intriguing that in a model of $m^2\phi^2$ chaotic inflation the one-loop correction to the two-point correlation function of inflaton fluctuations can be as large as $70\%$ for $N_*=45$. 

This seems to imply that the one-loop effects might have important consequences, when the cosmological data reaches the level where they in principle can rule out the $m^2\phi^2$ model of inflation. In fact, we are intrigued by the fact that cosmological data in the near future will reach the level where one appears to be able to discriminate between a pure low-energy effective $m^2\phi^2$ model of inflation with only the minimal total number of e-foldings, and the chaotic $m^2\phi^2$ model of inflation with a huge total number of e-foldings. This possibility deserves further studies.

However, one should be aware of some shortcomings in our treatment of one-loop corrections so far. For practical reasons, we have only calculated the one-loop correction to the scalar power-spectrum of inflaton fluctuations, while we have not yet been in the position to calculate the correction to the tensor spectrum. The calculation of the one-loop corrections to the scalar power-spectrum already required  us to calculate the fourth order action of scalar perturbations. If we desired to calculate the one-loop correction to the tensor power-spectrum, we would also need the fourth order action for the tensor modes, which has not yet been calculated. 

In addition, one should note that in the case of $m^2\phi^2$ chaotic inflation the one-loop effects can be so large, that the perturbative approach is on the verge of breaking down. This indicates that we in principle also should include two-loop effects in a precise treatment. However, this would require us to calculate the action of scalar perturbations beyond fourth order.

%%%%%%%%%%%%%%%%%%%%%%%%%%%%%%%%%%%%%%%%%%%%%%%%%%%%%%
%%%%%%%%%%%%%%%%%%%%%%%%%%%%%%%%%%%%%%%%%%%%%%%%%%%%%%
\overskrift{Acknowledgments}

%%%%%%%%%%%%%%%%%%%%%%%%%%%%%%%%%%%%%%%%%%%%%%%%%%%%%%
%%%%%%%%%%%%%%%%%%%%%%%%%%%%%%%%%%%%%%%%%%%%%%%%%%%%%%
\noindent I would like to thank Kari Enqvist, Steen Hannestad, Nemanja Kaloper, Shinsuke Kawai, James Lidsey, David Lyth, Sami Nurmi, David Seery and Filippo Vernizzi for interesting and motivating discussions. I would also like to thank Kari Enqvist and Helsinki Institute of Physics for the hospitality during the completion of parts of this work, and David Lyth and the Department of Physics at Lancaster University for the hospitality during the  workshop {\it Non-gaussianity from Inflation, June 2006}.

%%%%%%%%%%%%%%%%%%%%%%%%%%%%%%%%%%%%%%%%%%%%%%%%%%%%%%
%%%%%%%%%%%%%%%%%%%%%%%%%%%%%%%%%%%%%%%%%%%%%%%%%%%%%%


\begin{thebibliography}{99}

%\cite{Spergel:2006hy}
\bibitem{Spergel:2006hy}
  D.~N.~Spergel {\it et al.},
  %``Wilkinson Microwave Anisotropy Probe (WMAP) three year results:
  %Implications for cosmology,''
  arXiv:astro-ph/0603449.
  %%CITATION = ASTRO-PH 0603449;%%

%\cite{Kinney:2006qm}
\bibitem{Kinney:2006qm}
  W.~H.~Kinney, E.~W.~Kolb, A.~Melchiorri and A.~Riotto,
  %``Inflation model constraints from the Wilkinson microwave anisotropy  probe
  %three-year data,''
  Phys.\ Rev.\ D {\bf 74} (2006) 023502
  [arXiv:astro-ph/0605338].
  %%CITATION = ASTRO-PH 0605338;%%

%\cite{Tegmark:2006az}
\bibitem{Tegmark:2006az}
  M.~Tegmark {\it et al.},
  %``Cosmological Constraints from the SDSS Luminous Red Galaxies,''
  arXiv:astro-ph/0608632.
  %%CITATION = ASTRO-PH 0608632;%%


%\cite{Hamann:2006pf}
\bibitem{Hamann:2006pf}
  J.~Hamann, S.~Hannestad, M.~S.~Sloth and Y.~Y.~Y.~Wong,
  %``How robust are inflation model and dark matter constraints from
  %cosmological data?,''
  arXiv:astro-ph/0611582.
  %%CITATION = ASTRO-PH 0611582;%%


%\cite{Sloth:2006az}
\bibitem{Sloth:2006az}
  M.~S.~Sloth,
  %``On the one loop corrections to inflation and the CMB anisotropies,''
  Nucl.\ Phys.\ B {\bf 748} (2006) 149
  [arXiv:astro-ph/0604488].
  %%CITATION = ASTRO-PH 0604488;%%
  
 
%\cite{Linde:1983gd}
\bibitem{Linde:1983gd}
  A.~D.~Linde,
  %``Chaotic Inflation,''
  Phys.\ Lett.\ B {\bf 129} (1983) 177.
  %%CITATION = PHLTA,B129,177;%%

%%%%%%%%%%%%%%%%%%%%%%%%%%%%%%%%%%%%%%%%%%%%%%%%%%

%\cite{Abramo:1998hi,Abramo:2001dc,Abramo:2001dd,Mukhanov:1996ak,Abramo:1997hu,Afshordi:2000nr,Finelli:2001bn,Finelli:2003bp,Losic:2005vg,Martineau:2005aa,Wu:2006xp}


%\cite{Mukhanov:1996ak}
\bibitem{Mukhanov:1996ak}
  V.~F.~Mukhanov, L.~R.~W.~Abramo and R.~H.~Brandenberger,
  %``On the back reaction problem for gravitational perturbations,''
  Phys.\ Rev.\ Lett.\  {\bf 78}, 1624 (1997)
  [arXiv:gr-qc/9609026].
  %%CITATION = GR-QC 9609026;%%

%\cite{Abramo:1997hu}
\bibitem{Abramo:1997hu}
  L.~R.~W.~Abramo, R.~H.~Brandenberger and V.~F.~Mukhanov,
  %``The energy-momentum tensor for cosmological perturbations,''
  Phys.\ Rev.\ D {\bf 56}, 3248 (1997)
  [arXiv:gr-qc/9704037].
  %%CITATION = GR-QC 9704037;%%


%\cite{Abramo:1998hi}
\bibitem{Abramo:1998hi}
  L.~R.~W.~Abramo and R.~P.~Woodard,
  %``One loop back reaction on chaotic inflation,''
  Phys.\ Rev.\ D {\bf 60}, 044010 (1999)
  [arXiv:astro-ph/9811430].
  %%CITATION = ASTRO-PH 9811430;%%


%\cite{Afshordi:2000nr}
\bibitem{Afshordi:2000nr}
  N.~Afshordi and R.~H.~Brandenberger,
  %``Super-Hubble nonlinear perturbations during inflation,''
  Phys.\ Rev.\ D {\bf 63}, 123505 (2001)
  [arXiv:gr-qc/0011075].
  %%CITATION = GR-QC 0011075;%%



%\cite{Abramo:2001dc}
\bibitem{Abramo:2001dc}
  L.~R.~Abramo and R.~P.~Woodard,
  %``No one loop back-reaction in chaotic inflation,''
  Phys.\ Rev.\ D {\bf 65}, 063515 (2002)
  [arXiv:astro-ph/0109272].
  %%CITATION = ASTRO-PH 0109272;%%

%\cite{Abramo:2001dd}
\bibitem{Abramo:2001dd}
  L.~R.~Abramo and R.~P.~Woodard,
  %``Back-reaction is for real,''
  Phys.\ Rev.\ D {\bf 65}, 063516 (2002)
  [arXiv:astro-ph/0109273].
  %%CITATION = ASTRO-PH 0109273;%%

%\cite{Finelli:2001bn}
\bibitem{Finelli:2001bn}
  F.~Finelli, G.~Marozzi, G.~P.~Vacca and G.~Venturi,
% ``Energy-momentum tensor of field fluctuations in massive chaotic
  %inflation,''
  Phys.\ Rev.\ D {\bf 65}, 103521 (2002)
  [arXiv:gr-qc/0111035];
  %%CITATION = GR-QC 0111035;%%

%\cite{Finelli:2003bp}
\bibitem{Finelli:2003bp}
  F.~Finelli, G.~Marozzi, G.~P.~Vacca and G.~Venturi,
  %``Energy-momentum tensor of cosmological fluctuations during inflation,''
  Phys.\ Rev.\ D {\bf 69}, 123508 (2004)
  [arXiv:gr-qc/0310086];
  %%CITATION = GR-QC 0310086;%%


%\cite{Losic:2005vg}
\bibitem{Losic:2005vg}
  B.~Losic and W.~G.~Unruh,
  %``Long-wavelength metric backreactions in slow-roll inflation,''
  Phys.\ Rev.\ D {\bf 72}, 123510 (2005)
  [arXiv:gr-qc/0510078].
  %%CITATION = GR-QC 0510078;%%

%\cite{Martineau:2005aa}
\bibitem{Martineau:2005aa}
  P.~Martineau and R.~H.~Brandenberger,
  % ``The effects of gravitational back-reaction on cosmological
  %perturbations,''
  Phys.\ Rev.\ D {\bf 72}, 023507 (2005)
  [arXiv:astro-ph/0505236].
  %%CITATION = ASTRO-PH 0505236;%%

%\cite{Wu:2006xp}
\bibitem{Wu:2006xp}
  C.~H.~Wu, K.~W.~Ng, W.~Lee, D.~S.~Lee and Y.~Y.~Charng,
  %``Quantum noise and a low cosmic microwave background quadrupole,''
  arXiv:astro-ph/0604292.
  %%CITATION = ASTRO-PH 0604292;%%
  
  %%%%%%%%%%%%%%%%%%%%%%%%%%%%%%%%%%%%%%%%%%%%%%%%%%
  
%\cite{Unruh:1998ic,Geshnizjani:2002wp,Geshnizjani:2003cn}
  
  %\cite{Unruh:1998ic}
\bibitem{Unruh:1998ic}
  W.~Unruh,
  %``Cosmological long wavelength perturbations,''
  arXiv:astro-ph/9802323.
  %%CITATION = ASTRO-PH 9802323;%%
  
  %\cite{Geshnizjani:2002wp}
\bibitem{Geshnizjani:2002wp}
  G.~Geshnizjani and R.~Brandenberger,
  %``Back reaction and local cosmological expansion rate,''
  Phys.\ Rev.\ D {\bf 66} (2002) 123507
  [arXiv:gr-qc/0204074].
  %%CITATION = GR-QC 0204074;%%
  
  %\cite{Geshnizjani:2003cn}
\bibitem{Geshnizjani:2003cn}
  G.~Geshnizjani and R.~Brandenberger,
  %``Back reaction of perturbations in two scalar field inflationary models,''
  JCAP {\bf 0504} (2005) 006
  [arXiv:hep-th/0310265].
  %%CITATION = HEP-TH 0310265;%%
  

%%%%%%%%%%%%%%%%%%%%%%%%%%%%%%%%%%%%%%%%%%%%%%%%%%%

%\cite{Dimopoulos:2001wx,Prokopec:2004au,Prokopec:2002jn,Prokopec:2002uw,Prokopec:2003bx,Kahya:2006ui}

%\cite{Dimopoulos:2001wx}
\bibitem{Dimopoulos:2001wx}
  K.~Dimopoulos, T.~Prokopec, O.~Tornkvist and A.~C.~Davis,
  %``Natural magnetogenesis from inflation,''
  Phys.\ Rev.\ D {\bf 65} (2002) 063505
  [arXiv:astro-ph/0108093].
  %%CITATION = ASTRO-PH 0108093;%%

%\cite{Prokopec:2004au}
\bibitem{Prokopec:2004au}
  T.~Prokopec and E.~Puchwein,
  %``Nearly minimal magnetogenesis,''
  Phys.\ Rev.\ D {\bf 70} (2004) 043004
  [arXiv:astro-ph/0403335].
  %%CITATION = ASTRO-PH 0403335;%%


%\cite{Prokopec:2002jn}
\bibitem{Prokopec:2002jn}
  T.~Prokopec, O.~Tornkvist and R.~P.~Woodard,
  %``Photon mass from inflation,''
  Phys.\ Rev.\ Lett.\  {\bf 89} (2002) 101301
  [arXiv:astro-ph/0205331].
  %%CITATION = ASTRO-PH 0205331;%%

%\cite{Prokopec:2002uw}
\bibitem{Prokopec:2002uw}
  T.~Prokopec, O.~Tornkvist and R.~P.~Woodard,
  %``One loop vacuum polarization in a locally de Sitter background,''
  Annals Phys.\  {\bf 303} (2003) 251
  [arXiv:gr-qc/0205130].
  %%CITATION = GR-QC 0205130;%%

%\cite{Prokopec:2003bx}
\bibitem{Prokopec:2003bx}
  T.~Prokopec and R.~P.~Woodard,
  %``Vacuum polarization and photon mass in inflation,''
  Am.\ J.\ Phys.\  {\bf 72} (2004) 60
  [arXiv:astro-ph/0303358].
  %%CITATION = ASTRO-PH 0303358;%%

%\cite{Kahya:2006ui}
\bibitem{Kahya:2006ui}
  E.~O.~Kahya and R.~P.~Woodard,
  %``One loop corrected mode functions for SQED during inflation,''
  Phys.\ Rev.\ D {\bf 74} (2006) 084012
  [arXiv:gr-qc/0608049].
  %%CITATION = GR-QC 0608049;%%


%%%%%%%%%%%%%%%%%%%%%%%%%%%%%%%%%%%%%%%%%%%%%%%%%%%

%\cite{Linde:1986fd}
\bibitem{Linde:1986fd}
  A.~D.~Linde,
  %``Eternally Existing Selfreproducing Chaotic Inflationary Universe,''
  Phys.\ Lett.\ B {\bf 175}, 395 (1986).
  %%CITATION = PHLTA,B175,395;%%

%\cite{Goncharov:1987ir}
\bibitem{Goncharov:1987ir}
  A.~S.~Goncharov, A.~D.~Linde and V.~F.~Mukhanov,
  %``The Global Structure Of The Inflationary Universe,''
  Int.\ J.\ Mod.\ Phys.\ A {\bf 2}, 561 (1987).
  %%CITATION = IMPAE,A2,561;%%

%%%%%%%%%%%%%%%%%%%%%%%%%%%%%%%%%%%%%%%%%%%%%%%%%%%


 %\cite{Seery:2006vu}
\bibitem{Seery:2006vu}
  D.~Seery, J.~E.~Lidsey and M.~S.~Sloth,
  %``The inflationary trispectrum,''
  arXiv:astro-ph/0610210.
  %%CITATION = ASTRO-PH 0610210;%%

%\cite{Weinberg:2005vy}
\bibitem{Weinberg:2005vy}
  S.~Weinberg,
  %``Quantum contributions to cosmological correlations,''
  Phys.\ Rev.\ D {\bf 72}, 043514 (2005).
  %%CITATION = PHRVA,D72,043514;%%

%\cite{Weinberg:2006ac}
\bibitem{Weinberg:2006ac}
  S.~Weinberg,
  %``Quantum contributions to cosmological correlations. II: Can these
  %corrections become large?,''
  Phys.\ Rev.\ D {\bf 74} (2006) 023508
  [arXiv:hep-th/0605244].
  %%CITATION = HEP-TH 0605244;%%




%\cite{Maldacena:2002vr}
\bibitem{Maldacena:2002vr}
  J.~Maldacena,
  %``Non-Gaussian features of primordial fluctuations in single field
  %inflationary models,''
  JHEP {\bf 0305} (2003) 013
  [arXiv:astro-ph/0210603].
  %%CITATION = ASTRO-PH 0210603;%%




%\cite{Arnowitt:1962hi}
\bibitem{Arnowitt:1962hi}
  R.~Arnowitt, S.~Deser and C.~W.~Misner,
  %``The Dynamics Of General Relativity,''
  arXiv:gr-qc/0405109.
  %%CITATION = GR-QC 0405109;%%


%\cite{Creminelli:2003iq}
\bibitem{Creminelli:2003iq}
  P.~Creminelli,
  %``On non-gaussianities in single-field inflation,''
  JCAP {\bf 0310} (2003) 003
  [arXiv:astro-ph/0306122].
  %%CITATION = ASTRO-PH 0306122;%%

%\cite{Seery:2005wm}
\bibitem{Seery:2005wm}
  D.~Seery and J.~E.~Lidsey,
  %``Primordial non-gaussianities in single field inflation,''
  JCAP {\bf 0506} (2005) 003
  [arXiv:astro-ph/0503692].
  %%CITATION = ASTRO-PH 0503692;%%

%\cite{Seery:2005gb}
\bibitem{Seery:2005gb}
  D.~Seery and J.~E.~Lidsey,
  %``Primordial non-gaussianities from multiple-field inflation,''
  JCAP {\bf 0509} (2005) 011
  [arXiv:astro-ph/0506056].
  %%CITATION = ASTRO-PH 0506056;%%

%\cite{Schwinger:1960qe}
\bibitem{Schwinger:1960qe}
  J.~S.~Schwinger,
  %``Brownian Motion Of A Quantum Oscillator,''
  J.\ Math.\ Phys.\  {\bf 2}, 407 (1961).
  %%CITATION = JMAPA,2,407;%%

%\cite{Keldysh:1964ud}
\bibitem{Keldysh:1964ud}
  L.~V.~Keldysh,
  %``Diagram Technique For Nonequilibrium Processes,''
  Zh.\ Eksp.\ Teor.\ Fiz.\  {\bf 47}, 1515 (1964)
  [Sov.\ Phys.\ JETP {\bf 20}, 1018 (1965)].
  %%CITATION = ZETFA,47,1515;%%

%%%%%%%%%%%%%%%%%%%%%%%%%%%%%%%%%%%%%%%%%%%

%\cite{Calzetta:1986ey,Calzetta:1986cq,Boyanovsky:1992vi,Boyanovsky:1996rw,Boyanovsky:1997xt,Boyanovsky:2000hs}


%\cite{Calzetta:1986ey}
\bibitem{Calzetta:1986ey}
  E.~Calzetta and B.~L.~Hu,
  %``CLOSED TIME PATH FUNCTIONAL FORMALISM IN CURVED SPACE-TIME: APPLICATION TO
  %COSMOLOGICAL BACK REACTION PROBLEMS,''
  Phys.\ Rev.\ D {\bf 35} (1987) 495.
  %%CITATION = PHRVA,D35,495;%%

%\cite{Calzetta:1986cq}
\bibitem{Calzetta:1986cq}
  E.~Calzetta and B.~L.~Hu,
  %``NONEQUILIBRIUM QUANTUM FIELDS: CLOSED TIME PATH EFFECTIVE ACTION, WIGNER
  %FUNCTION AND BOLTZMANN EQUATION,''
  Phys.\ Rev.\ D {\bf 37} (1988) 2878.
  %%CITATION = PHRVA,D37,2878;%%

%\cite{Boyanovsky:1992vi}
\bibitem{Boyanovsky:1992vi}
  D.~Boyanovsky and H.~J.~de Vega,
  %``Quantum rolling down out-of-equilibrium,''
  Phys.\ Rev.\ D {\bf 47} (1993) 2343
  [arXiv:hep-th/9211044].
  %%CITATION = HEP-TH 9211044;%%

%\cite{Boyanovsky:1996rw}
\bibitem{Boyanovsky:1996rw}
  D.~Boyanovsky, D.~Cormier, H.~J.~de Vega and R.~Holman,
  %``Out of equilibrium dynamics of an inflationary phase transition,''
  Phys.\ Rev.\ D {\bf 55} (1997) 3373
  [arXiv:hep-ph/9610396].
  %%CITATION = HEP-PH 9610396;%%

%\cite{Boyanovsky:1997xt}
\bibitem{Boyanovsky:1997xt}
  D.~Boyanovsky, D.~Cormier, H.~J.~de Vega, R.~Holman and S.~P.~Kumar,
  %``Non-perturbative quantum dynamics of a new inflation model,''
  Phys.\ Rev.\ D {\bf 57} (1998) 2166
  [arXiv:hep-ph/9709232].
  %%CITATION = HEP-PH 9709232;%%

%\cite{Boyanovsky:2000hs}
\bibitem{Boyanovsky:2000hs}
  D.~Boyanovsky and H.~J.~de Vega,
  %``Out of equilibrium fields in selfconsistent inflationary dynamics.  Density
  %fluctuations,''
  arXiv:astro-ph/0006446.
  %%CITATION = ASTRO-PH 0006446;%%



%%%%%%%%%%%%%%%%%%%%%%%%%%%%%%%%%%%%%%%%%%%

%%%%%%%%%%%%%%%%%%%%%%%%%%%%%%%%%%%%%%%%%%%

%\cite{Boyanovsky:2004gq,Boyanovsky:2004ph,Boyanovsky:2005sh,Boyanovsky:2005px,Boyanovsky:2006kg}

%\cite{Boyanovsky:2004gq}
\bibitem{Boyanovsky:2004gq}
  D.~Boyanovsky and H.~J.~de Vega,
  %``Particle decay in inflationary cosmology,''
  Phys.\ Rev.\ D {\bf 70} (2004) 063508
  [arXiv:astro-ph/0406287].
  %%CITATION = ASTRO-PH 0406287;%%

%\cite{Boyanovsky:2004ph}
\bibitem{Boyanovsky:2004ph}
  D.~Boyanovsky, H.~J.~de Vega and N.~G.~Sanchez,
  %``Particle decay during inflation: Self-decay of inflaton quantum
  %fluctuations during slow roll,''
  Phys.\ Rev.\ D {\bf 71} (2005) 023509
  [arXiv:astro-ph/0409406].
  %%CITATION = ASTRO-PH 0409406;%%

%\cite{Boyanovsky:2005sh}
\bibitem{Boyanovsky:2005sh}
  D.~Boyanovsky, H.~J.~de Vega and N.~G.~Sanchez,
  %``Quantum corrections to slow roll inflation and new scaling of  superhorizon
  %fluctuations,''
  Nucl.\ Phys.\ B {\bf 747} (2006) 25
  [arXiv:astro-ph/0503669].
  %%CITATION = ASTRO-PH 0503669;%%

%\cite{Boyanovsky:2005px}
\bibitem{Boyanovsky:2005px}
  D.~Boyanovsky, H.~J.~de Vega and N.~G.~Sanchez,
  %``Quantum corrections to the inflaton potential and the power spectra  from
  %superhorizon modes and trace anomalies,''
  Phys.\ Rev.\ D {\bf 72} (2005) 103006
  [arXiv:astro-ph/0507596].
  %%CITATION = ASTRO-PH 0507596;%%

%\cite{Boyanovsky:2006kg}
\bibitem{Boyanovsky:2006kg}
  D.~Boyanovsky, H.~J.~de Vega and N.~G.~Sanchez,
  %``Clarifying slow roll inflation and the quantum corrections to the
  %observable power spectra,''
  arXiv:astro-ph/0601132.
  %%CITATION = ASTRO-PH 0601132;%%


%%%%%%%%%%%%%%%%%%%%%%%%%%%%%%%%%%%%%%%%%%%








%%%%%%%%%%%%%%%%%%%%%%%%%%%%%%%%%%%%%%%%%%%
%\cite{Collins:2005nu}
\bibitem{Collins:2005nu}
  H.~Collins and R.~Holman,
  %``Renormalization of initial conditions and the trans-Planckian problem  of
  %inflation,''
  Phys.\ Rev.\ D {\bf 71} (2005) 085009
  [arXiv:hep-th/0501158].
  %%CITATION = HEP-TH 0501158;%%

%\cite{Matacz:1992tp}
\bibitem{Matacz:1992tp}
  A.~L.~Matacz,
  %``The Coherent state representation of quantum fluctuations in the early
  %universe,''
  Phys.\ Rev.\ D {\bf 49} (1994) 788
  [arXiv:gr-qc/9212008].
  %%CITATION = GR-QC 9212008;%%

%\cite{chisholm}
\bibitem{chisholm}
J.~S.~R.~Chisholm,
  %``Change of variables in quantum field theories,''
  Nucl.\ Phys.\  {\bf 26} (1961) 469.


%\cite{Kamefuchi:1961sb}
\bibitem{Kamefuchi:1961sb}
  S.~Kamefuchi, L.~O'Raifeartaigh and A.~Salam,
  %``Change of variables and equivalence theorems in quantum field theories,''
  Nucl.\ Phys.\  {\bf 28} (1961) 529.
  %%CITATION = NUPHA,28,529;%%
  
  %\cite{Enqvist:2006fs}
\bibitem{Enqvist:2006fs}
  K.~Enqvist, J.~Hogdahl, S.~Nurmi and F.~Vernizzi,
  %``A covariant generalization of cosmological perturbation theory,''
  arXiv:gr-qc/0611020.
  %%CITATION = GR-QC 0611020;%%

\end{thebibliography}
\end{document}